\documentclass[10pt,conference]{IEEEtran}
\IEEEoverridecommandlockouts

\usepackage{flushend}

\usepackage{cite}
\def\BibTeX{{\rm B\kern-.05em{\sc i\kern-.025em b}\kern-.08em
    T\kern-.1667em\lower.7ex\hbox{E}\kern-.125emX}}

\newcommand\code[1]{{\tt \small #1}}

\usepackage{multirow}

\usepackage{tikz}

\usepackage{listings,pmboxdraw}
\usepackage{algorithm}
\usepackage{amsmath}
\usepackage[noend]{algpseudocode}
\usepackage[normalem]{ulem}

\usepackage{cancel}

\algblockdefx[Foreach]{Foreach}{EndForeach}[1]{\textbf{foreach} #1 \textbf{do}}{\textbf{end foreach}}

\makeatletter
\def\BState{\State\hskip-\ALG@thistlm}
\makeatother

\usepackage{amssymb}
\usepackage{pifont}

\lstdefinelanguage{JavaScript}{
  keywords={typeof, new, true, false, catch, function, return, null, catch, switch, var, if, in, while, do, else, case, break},
  keywordstyle=\color{blue}\bfseries,
  ndkeywords={class, export, boolean, throw, implements, import, this},
  ndkeywordstyle=\color{darkgray}\bfseries,
  identifierstyle=\color{black},
  sensitive=false,
  comment=[l]{//},
  morecomment=[s]{/*}{*/},
  stringstyle=\color{red}\ttfamily,
  morestring=[b]',
  morestring=[b]",
  morestring=[b]`,
  morecomment=[f][\lstbg{red!20}]-,
  morecomment=[f][\lstbg{green!20}]+,
  escapeinside=||,
  numbers=none
}

\lstdefinelanguage{JavaScriptNumbers}{
  language = JavaScript,
  numbers=left,
  xleftmargin=2em,
  framexleftmargin=1.5em
}

\usepackage{xcolor}
\usepackage{booktabs}

\colorlet{punct}{red!60!black}
\definecolor{background}{HTML}{EEEEEE}
\definecolor{delim}{RGB}{20,105,176}
\colorlet{numb}{magenta!60!black}

\lstdefinelanguage{special}{
  basicstyle=\ttfamily,
  columns=fullflexible,
  keepspaces,
  literate=
  {┐}{\textSFiii}1%
  {└}{\textSFii}1%
  {┴}{\textSFvii}1%
  {┬}{\textSFvi}1%
  {├}{\textSFviii}{1}%
  {─}{\textSFx}1%
  {┼}{\textSFv}1,
}
\lstdefinelanguage{json}{
    basicstyle=\small\ttfamily,
    numbers=left,
    numberstyle=\scriptsize,
    stepnumber=1,
    numbersep=8pt,
    showstringspaces=false,
    breaklines=true,
    frame=lines,
    literate=
     *{0}{{{\color{numb}0}}}{1}
      {1}{{{\color{numb}1}}}{1}
      {2}{{{\color{numb}2}}}{1}
      {3}{{{\color{numb}3}}}{1}
      {4}{{{\color{numb}4}}}{1}
      {5}{{{\color{numb}5}}}{1}
      {6}{{{\color{numb}6}}}{1}
      {7}{{{\color{numb}7}}}{1}
      {8}{{{\color{numb}8}}}{1}
      {9}{{{\color{numb}9}}}{1}
      {:}{{{\color{punct}{:}}}}{1}
      {,}{{{\color{punct}{,}}}}{1}
      {\{}{{{\color{delim}{\{}}}}{1}
      {\}}{{{\color{delim}{\}}}}}{1}
      {[}{{{\color{delim}{[}}}}{1}
      {]}{{{\color{delim}{]}}}}{1},
}

\usepackage{amsthm}

\usepackage{array}
\newcolumntype{P}[1]{>{\centering\arraybackslash}p{#1}}
\newcolumntype{M}[1]{>{\centering\arraybackslash}m{#1}}

\newtheoremstyle{sltheorem}
{0.5}                
{0}                
{}        
{\parindent}                
{\bfseries}       
{.}               
{ }               
{}                
\theoremstyle{sltheorem}

\newcommand\fwnoformat{{TESA}}
\newcommand\fw{{\sf \fwnoformat{}}}

\newcommand\npmreg{npm registry}
\usepackage{xspace}

\usepackage{paralist}

\newcommand\TOTAL[0]{500}
\newcommand\mv[0]{$n@v$}

\author{\IEEEauthorblockN{Haiyang Sun\IEEEauthorrefmark{1},
Andrea Ros\`a\IEEEauthorrefmark{1},
Daniele Bonetta\IEEEauthorrefmark{2}, and
Walter Binder\IEEEauthorrefmark{1}}
\IEEEauthorblockA{\IEEEauthorrefmark{1}Universit\`a della Svizzera Italiana, Faculty of Informatics, Lugano, Switzerland\\ Email: \{haiyang.sun, andrea.rosa, walter.binder\}@usi.ch}
\IEEEauthorblockA{\IEEEauthorrefmark{2}Oracle Labs, Netherlands\\
Email: daniele.bonetta@oracle.com}
}

\begin{document}

\title{Automatically Assessing and Extending Code Coverage for NPM Packages}

\maketitle


\begin{abstract}
Typical Node.js applications extensively rely on packages hosted
in the npm registry. As such packages may be used by thousands
of other packages or applications, it is important to assess their
code coverage. Moreover, increasing code coverage may help detect
previously unknown issues. In this paper, we introduce
\fw{}, a new tool that automatically assembles a test suite for any
package in the npm registry. The test suite includes 1) tests written
for the target package and usually hosted in its development repository, 
and 2) tests selected from dependent packages. The former
tests allow assessing the code coverage of the target package, while
the latter ones can increase code coverage by exploiting third-party
tests that also exercise code in the target package. We use \fw{} to
assess the code coverage of 500 popular npm packages. Then, we
demonstrate that \fw{} can significantly increase code coverage by
including tests from dependent packages. Finally, we show that the
test suites assembled by \fw{} increase the effectiveness of existing
dynamic program analyses to identify performance issues that are
not detectable when only executing the developer's tests.
\end{abstract}



\section{Introduction}\label{sec:introduction}
Node.js~\cite{5617064} has become one of the most popular runtimes for executing server-side JavaScript programs.
Node.js applications typically resort to hundreds of publicly available third-party \emph{packages} hosted in the \emph{Node package manager (npm) registry}~\cite{npm}, a large repository containing over a million of ready-to-use publicly available third-party libraries. 
While the availability of a large number of packages is appealing and eases the development of complex software, it is fundamental that such packages have high quality, to reduce the chances that a buggy package affects potentially many other packages and applications~\cite{10.1145/3377811.3380442,10.1145/3377811.3380426}.
Unfortunately, the lack of uniform code-quality standards exposes the npm ecosystem to bugs and vulnerabilities, whose negative impact increases with the popularity of the affected package~\cite{zimmermann2019small,10.1145/2901739.2901743,npmnews1,npmnews2,npmnews3,6470829,survey2017}.

As testing is important to ensure the quality of a program and one important aspect of the test quality is code coverage, having a test suite with good code coverage is fundamental.
A recent study by Fard et al.~\cite{7927978} investigates the tests of 86 packages from the npm registry and finds that many of them suffer from broken test configuration or poor coverage. 
Due to the lack of an automatic tool to test and measure the code coverage, their study was done manually.
Such an approach does not scale to the large number of available npm packages.

While the availability of a fully automated tool for testing and evaluating the code coverage of popular npm packages is crucial, developing such a tool is challenging.
An automatic approach needs to identify configuration failures in building and running the existing tests, which are very common~\cite{7927978}.
In addition, measuring the coverage of the tests shipped in a package's release in the \npmreg{} is insufficient in general. 
Many packages (more than 90\% of those analyzed in this paper) are intentionally deployed in the \npmreg{} without any testing code, so as to decrease the release size.
Hence, to find tests of a package of a specific version, one needs to consider also the package's development repository.
Moreover, as the repository may hold thousands of revisions (in terms of commits, tags, or releases), each of them often containing significantly different package code as well as different tests, an automatic approach needs to choose the revisions with testing code compatible with the version of the package in the \npmreg{}.

In this paper, we tackle these issues with \fw{}, a novel tool that enables automatic testing and coverage measurement for npm packages.
\fw{} can not only run and measure the code coverage for the \emph{original tests} for the target package located in the package's development repository, but also compose an additional test suite of \emph{dependent tests}
from third-party packages, treating the tests of third-party packages that rely on the target package (called \emph{dependent packages} in this paper, or \emph{dependents} for short) as tests for the target package itself.
While the former tests (written by the maintainers of the target package) allow testing implementation details of the target package, the latter (written by the dependent packages' developers) serve as black-box tests for testing (parts of) the public functionalities of the target package.
To the best of our knowledge, \fw{} is the first framework for the automatic assembly of test suites fully leveraging the abundance of code available in the \npmreg{}. 
\fw{} also provides a compaction algorithm that produces an optimal compacted test suite, i.e., containing only those tests necessary to maximize code coverage while minimizing the total test execution time. Our evaluation results 
show that \fw{} can increase the number of successful tests by fixing common misconfigurations automatically,
and that the test suite assembled by \fw{} can significantly improve the code coverage for many npm packages. 

Moreover, in addition to extending code coverage, \fw{} enables the automatic execution of arbitrary dynamic program analyses (DPAs) on the testing code (including both original and dependent tests) of any set of npm~packages, allowing the automated execution of large-scale DPAs ``in the wild''. 
Thanks to \fw{}, we run a state-of-the-art DPA for identifying performance problems in npm~packages. 
Our results demonstrate that dependent tests allow finding performance problems that cannot be detected with the original tests of a package, further confirming the benefits of extending code coverage by exploiting tests written for dependent packages.

To summarize, our work makes the following contributions. 
We present \fw{}, a novel tool for the automatic assembly of test suites for any package in the npm registry, including original tests from a package's development repository and dependent tests written for dependent packages (Section~\ref{sec:methodology}).
We use \fw{} to assess the code coverage of \TOTAL{} popular packages, and we show that \fw{} can significantly increase code coverage by including tests from dependent packages.
Moreover, we demonstrate that executing DPAs on the test suites assembled by \fw{} 
enables the identification of performance issues that are not detectable when only executing the original tests (Section~\ref{sec:eval}). 
We complement the paper with 
a discussion on the lessons learned while developing \fw{} and conducting the evaluation  (Section~\ref{sec:lessons}), 
an outline on important aspects of our approach (Section~\ref{sec:discussion}), and an overview of related work (Section~\ref{sec:rw}).
Finally, we draw our conclusions in Section~\ref{sec:conclusion}.

\section{\fwnoformat{}}\label{sec:methodology}

In this section, we introduce the approach used by \fw{} to assemble a test suite for a given npm package.

\subsection{Test Inclusion Criteria}\label{sec:cov}

\fw{} treats tests at the granularity of packages; henceforth, we consider as a \emph{single test} all the code that executes when invoking \code{npm test}, after package installation with \code{npm install}.
In this paper, we denote a package with name $n$ and version $v$ as $n@v$.  
For a test to be considered for inclusion in the test suite for \mv, a test (either original or dependent) has to satisfy two requirements:
\begin{compactitem}
    \item
        The test has to succeed.
    \item
        The test has to cover some code in \mv{}.
\end{compactitem}

As a successful \code{npm install} or \code{npm test} command always terminates with exit code zero, \fw{} checks the exit code to determine whether installation or test execution has succeeded or failed.
To determine whether a test covers some code in \mv, \fw{} measures code coverage using three different metrics, i.e., statement, function, and branch coverage, which are computed by \code{istanbul.js}.\footnote{\code{\footnotesize istanbul.js}~\cite{istanbuljs} is a popular tool for measuring code coverage in Node.js applications.} 
In principle, our approach is not limited to such test-coverage metrics and can be extended to use other metrics. 
The code coverage of a set of tests $ts$ for a package \mv{} is denoted as $Cov_{kind}(n@v, ts)$ ($kind$ denotes the program elements for which the code coverage is computed: statement, function, or branch) and is defined as follows:

\begin{equation}
    \footnotesize
    \begin{gathered}
        Total_{kind}(n@v) \gets\\ \{~id=(f,loc) \mid~\exists~\text{a}~kind\text{-element at}~loc~\text{in file}~f~\text{of}~n@v~\}
\\[3mm]
        Covered_{kind}(n@v, t \in ts) \gets \\ \{~id \in Total_{kind}(n@v) \mid t~\text{covers}~$id$~\}
\\[3mm]
        Cov_{kind}(n@v, ts) \gets \frac{|\bigcup\limits_{t \in ts} Covered_{kind}(n@v, t)|}{|Total_{kind}(n@v)|}
    \end{gathered}
    \label{eq:1}
\end{equation}

As defined in Formula~\ref{eq:1}, %
every program element of interest (statement, function, branch) is identified with a unique ID consisting of the enclosing source-code file (distinguished by the relative path of the file in the package) and the location of the program element in the file (i.e.,  the starting position of the element in the file).

When assembling the test suite $TS(n@v)$ for a package \mv{}, discovered tests~$t$ are included as follows: 
\begin{equation}
    \footnotesize
    \begin{gathered}
        TS(n@v) \gets \\ \{~t \mid~t~\text{succeeds}~~\land~~\exists~kind~\text{with}~Cov_{kind}(n@v, \{t\}) > 0~\}
    \end{gathered}
\end{equation}

\subsection{Finding Original Tests}\label{sec:findoriginaltests}
Here, we describe how \fw{} locates original tests for \mv{}, provided by the developers of package $n$ in its development repository.\footnote{The URL of the development repository is specified in the {\tt package.json}~file of the package.}
The development repository may contain thousands of revisions (i.e., commits, tags, and releases).
While in general, there is a mapping between the code of \mv{} released in the \npmreg{} and a revision in the development repository containing the same code, the corresponding revision may not contain suitable tests. 
To locate suitable tests, a simple approach would be to run \emph{all} tests from \emph{all} the revisions in the development repository and to measure code coverage wrt.~\mv{}.
However, such an approach would be too costly, particularly when assembling test suites for a large set of packages. 

\fw{} employs an automatic procedure to locate tests compatible with \mv{} in the development repository. 
In particular, \fw{} considers only the revisions that closely match the code of \mv{}. 
Our approach is based on the following empirical observation on development repositories: 
most development repositories have one or more releases corresponding to package \mv{} as published in the \npmreg.
Moreover, especially in well-maintained packages, the release ID contains the version number of the package as released in the \npmreg.

Motivated by this observation, \fw{} selects those releases whose ID textually contains version $v$. 
With high probability, most of the source-code files in the selected revisions are the same as those of \mv{} in the \npmreg{}.
If there are tests in such revisions, they are more likely to have higher coverage for \mv{} compared to other revisions.

Unfortunately, some packages do not follow this naming practice, or the selected revisions may contain no tests. 
To handle such cases, \fw{} also looks for tests in the revision corresponding to the last commit before the release date of \mv{} in the \npmreg{},  
which is likely to contain code matching \mv{}.  

However, the selected revisions 
may still contain code files
that are different in content wrt.~those in the~\npmreg{} for  \mv{}. 
Statements, branches, or functions covered in these different files will not be counted
in the coverage measurement. 
\fw{} attempts to fix such differences 
 automatically, by replacing the different files with the corresponding ones from the \npmreg{}.

To summarize, \fw{} looks for original tests from the development repository by selecting the revisions that best match \mv{}, attempting to automatically fix 
minor differences found between the selected revisions and the package code in the~\npmreg{}. 

\begin{figure*}[t]
    \centering
    \begin{tabular}{M{0.3\textwidth}M{0.3\textwidth}M{0.3\textwidth}}
   \hspace{-0.04\textwidth}      \includegraphics[width=0.3\textwidth]{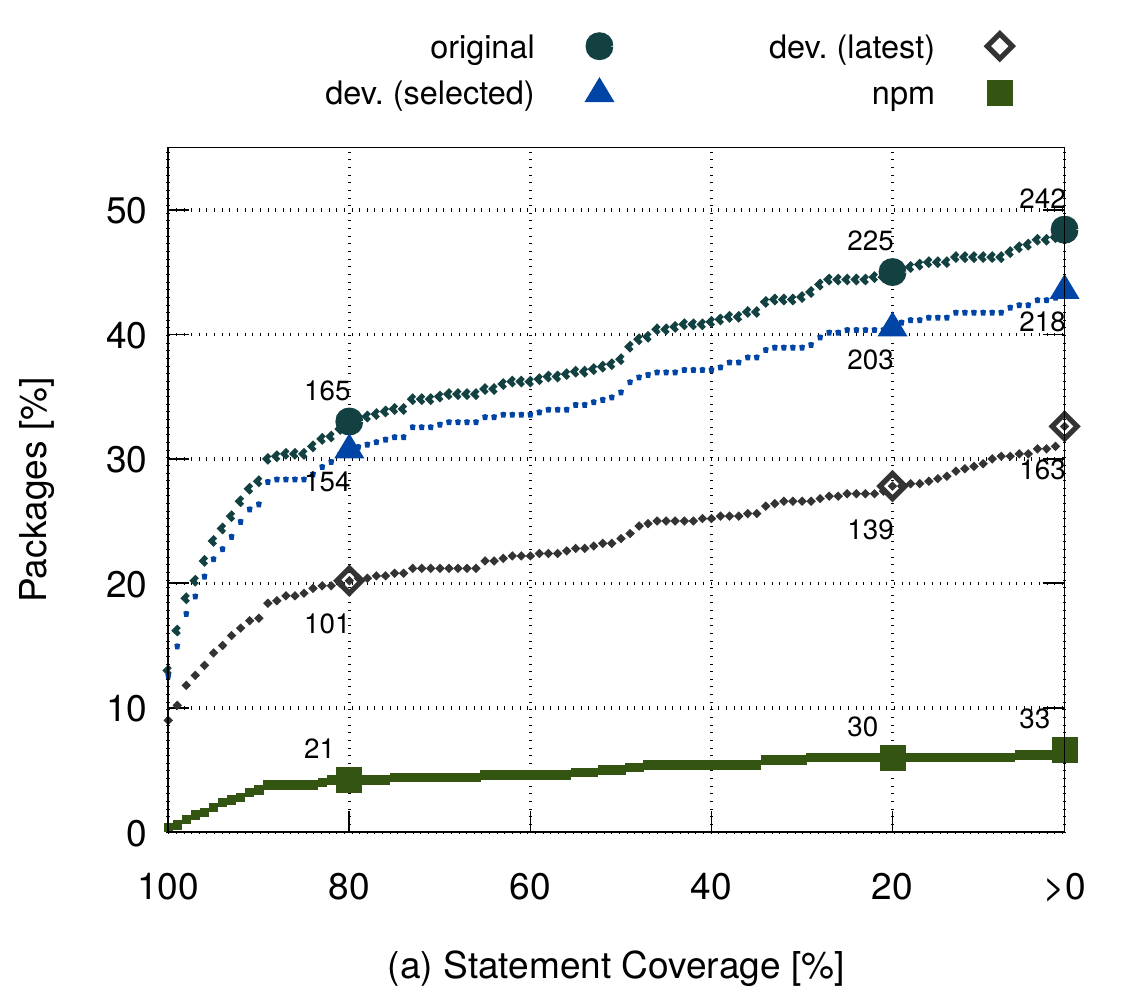} &
   \hspace{-0.04\textwidth}     \includegraphics[width=0.3\textwidth]{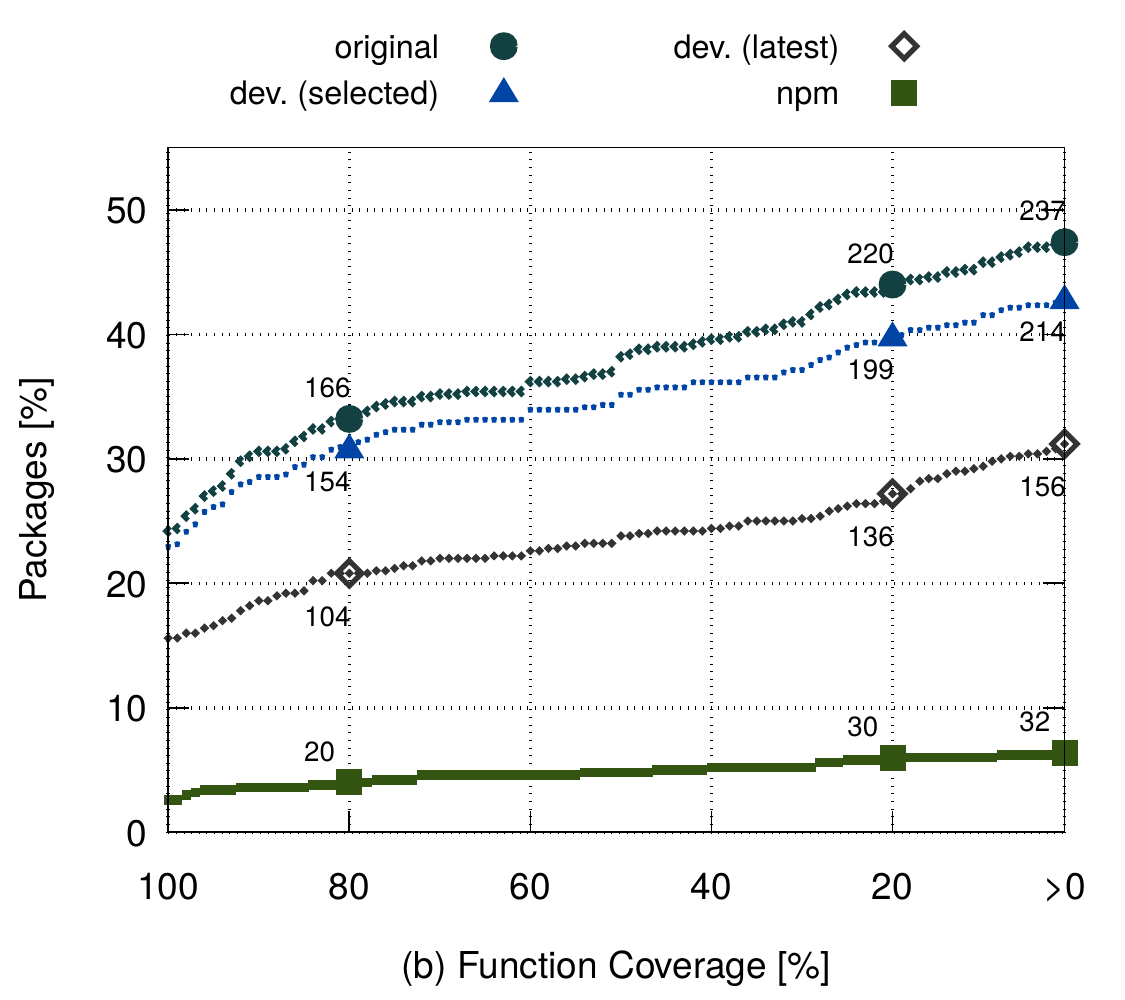} &
      \hspace{-0.04\textwidth}    \includegraphics[width=0.3\textwidth]{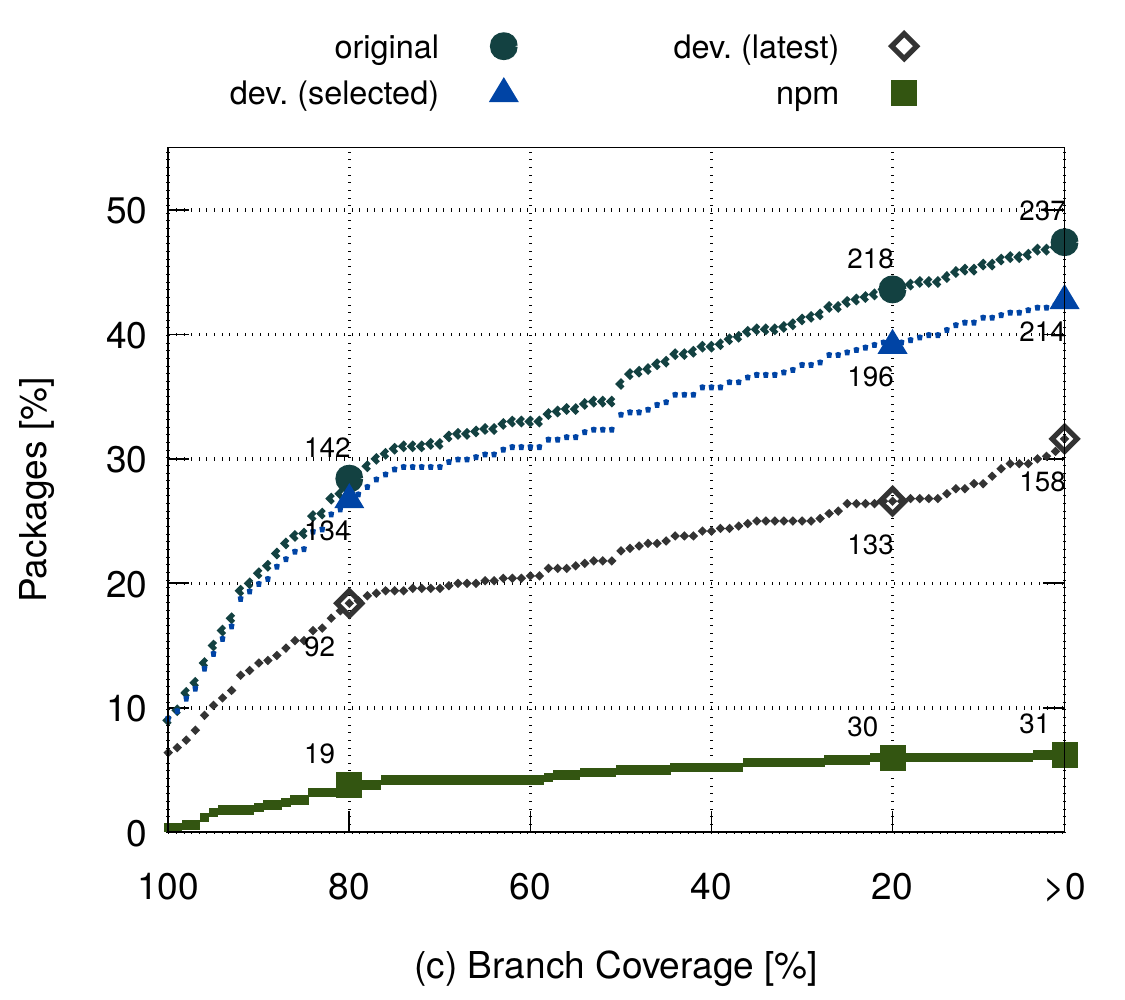} \\
    \end{tabular}
    \caption{Code coverage of the original tests found by \fw{}. Points in the curve represent the number of packages with coverage $\geq$ 80\%, $\geq$ 20\%, and $>$ 0\%, respectively.  The evaluation set is composed of \TOTAL{} packages.
    }
    \label{fig:ori-improve}
\end{figure*}

\subsection{Finding Dependent Tests}\label{sec:deptests}
Here,  we detail the approach used by \fw{} to look for dependent tests.
Such tests are important because they represent how \mv{} is used by dependent packages. 
The portions of \mv{} covered by dependent tests can significantly differ from those covered by the original tests.
As a result, including dependent tests can increase the coverage of the test suite. 
Moreover, dependent tests can be useful for a package developer to check whether a new update to his package may break its dependent packages.
However, finding high-quality dependents that contain tests for \mv{} is not trivial. 
While a popular package can have tens of thousands of dependents, many of them may not contain tests, or may only contain tests that do not cover \mv{}.

In a first step, \fw{} filters and sorts the available set of dependent packages. 
Our tool identifies and filters out dependents that are not compatible with the  version $v$ of \mv{}.
For example, if $v$ is \code{3.0.0} and a dependent package explicitly requires a version that is newer (e.g., \code{4.0.0}) or older (e.g., \code{2.0.0}) than \code{3.0.0}, 
such dependent cannot work with \mv{} and is therefore filtered out. 
In addition, \fw{} sorts the dependents according to their average daily downloads and 
evaluates them in descending order (starting with the most frequently downloaded dependent).
The search stops when a given maximum number of dependents is reached.

In a second step, \fw{} measures the coverage of the tests of the remaining dependents to determine whether they cover \mv{}.
The original tests of dependents can be located in the same way as the original tests of the target package (see Section~\ref{sec:findoriginaltests}).
In case a range version is specified, \fw{} forces the dependent test to download the exact version of \mv{} from the \npmreg{}. 
Without such an approach, a newer version than $v$ in the range (if any) would be automatically installed, resulting in code files that do not match \mv{}, hence leading to lower code coverage. 

\subsection{Compacting the Test Suite}\label{sec:compact}
The test suite $TS(n@v)$ resulting from the previous steps  (including both original and dependent tests) may be large and require a long time to run. 
Moreover, not all tests in $TS(n@v)$ are fundamental to improve the code coverage of \mv{}, as a test may cover only code portions that are already covered by other tests.
For these reasons, \fw{} uses a compaction algorithm to reduce the size of $TS(n@v)$.
Such a compacted test suite is particularly necessary in scenarios where developers are concerned about the time to run tests.

The compaction algorithm employed by \fw{} can be described as an optimization problem to find a subset $OPT(n@v) \subseteq TS(n@v)$ with the same code coverage, but requiring the shortest estimated execution time among all candidate solutions.
The specification of the optimization problem is given in Formula~\ref{eq:opt}, where
$\mathcal{P}(TS(n@v))$ is the power set of the test suite $TS(n@v)$, $CND(n@v)$ is the set of candidate solutions, and $OPT(n@v)$ is an optimal solution to the problem.

\begin{equation}
    \footnotesize
\label{eq:opt}
    \begin{gathered}
        \displaystyle
        CND(n@v) \gets \{~ts \in \mathcal{P}(TS(n@v)) \mid\\ \forall~kind: Cov_{kind}(n@v, ts) = Cov_{kind}(n@v, TS(n@v))~\}
        \\[3mm]
        \displaystyle
        \forall~ts \in \mathcal{P}(TS(n@v)): time(ts) \gets \sum_{t \in ts} time(t)
        \\[2mm]
        \displaystyle
    OPT(n@v) \gets \\ X \in CND(n@v) \mid \forall~ts \in CND(n@v): time(X) \leq time(ts)
    \end{gathered}
\end{equation}

In Formula~\ref{eq:opt}, $time(t)$ represents the time to execute a test $t$.
\fw{} measures time as the median of $N$ runs (user-defined value) of $t$ without any coverage measurement, to avoid any measurement perturbations caused by \code{istanbul.js}. 
\fw{} uses a general branch-and-bound~\cite{lawler1966branch} algorithm to find the optimal solution for the problem defined in Formula~\ref{eq:opt}.
The compaction rate of our algorithm for $TS(n@v)$ (i.e., the estimated speedup of running the tests in $OPT(n@v)$ vs.~the tests in $TS(n@v)$) can be computed as follows:

\begin{equation}
    \footnotesize
    CompactionRate(TS(n@v)) \gets \frac{time(TS(n@v))}{time(OPT(n@v))}
\end{equation}


\newcommand\ts[0]{evaluation set}

\section{Evaluation}\label{sec:eval}

In this section, we evaluate the test suites assembled by \fw{}. 
Our experiments target test suites assembled by our tool for an \ts{} including \TOTAL{} very popular and influential packages. 
For each package, we consider its last released version before January 1, 2020. 
The packages are chosen according to several popularity criteria, i.e., the number of downloads, GitHub stars, and the number of direct dependent packages.
More concretely, packages in the \ts{} must have at least 10K daily downloads on average in 2019  and more than 100~GitHub stars in their development repository. 
Moreover, the considered version of each package must have at least 20 other dependent packages. 
Overall, packages in the \ts{} have 673K daily downloads, 12K GitHub stars, and 611 dependents on average.

\subsection{Original Tests}\label{sec:eval_ori}
Here, we evaluate the coverage of the original tests for the packages in the \ts, automatically measured by \fw{}. 
We show the results in Figure~\ref{fig:ori-improve}, where each subfigure represents one of the coverage metrics used by \fw{}, i.e., statement (Figure~\ref{fig:ori-improve}a), function (Figure~\ref{fig:ori-improve}b), and branch coverage (Figure~\ref{fig:ori-improve}c).
Each graph shows 4 curves, each reporting the percentage of packages of the \ts{} (y-axis) whose tests satisfy a coverage threshold (x-axis).
Each curve considers different tests, as follows: 
\begin{compactitem}
    \item
        \emph{npm} -- considers only the tests found in the package release in the \npmreg.
    \item
        \emph{dev.~(latest)} -- considers only the tests located in the latest commit of the master branch of the development repository. Such revision represents the first place where developers usually search for tests. 
    \item
        \emph{dev.~(selected)} -- considers not only the latest revision, but also the ones that \fw{} determines to be related to the package version (i.e., releases matching the  version $v$ of the target module and the last commit before the release date of the package in the \npmreg{}).
    \item
        \emph{original} -- considers all original tests as specified in Section~\ref{sec:findoriginaltests} (i.e., after fixing files that are different from the module release). %
\end{compactitem}

The number of packages that have coverage $\geq$ 80\%, $\geq$ 20\%, and $>$ 0\% is shown nearby the respective points in the curves.

As the trends for the three coverage metrics are similar, we explain our results focusing only on statement coverage (Figure~\ref{fig:ori-improve}a).
The discussion below can be generalized also to function and branch coverage. 
From the \emph{npm} curve, we observe that only 4.2\% of the packages in the \ts{}  have statement coverage $\geq$ 80\%, only 6.0\% of the packages have coverage $\geq$ 20\%, and only 6.6\% of the packages contain tests with non-zero coverage. 
This result confirms that most packages do not have tests in their releases in the \npmreg, or have only a limited portion of them.

When looking at the \emph{dev.~(latest)} curve, the trend shows that looking for suitable tests considering only the latest commit in the development repository can already greatly improve code coverage.
For example, the number of packages with coverage $\geq$ 80\% increases by a factor of 4.8x, the number of packages with coverage $\geq$ 20\%  increases by a factor of 4.6x, while the number of packages with coverage $>$ 0\% increases by a factor of 4.9x. 
However, as indicated by the \emph{dev.~(selected)} curve, the approach used by \fw{} to select revisions matching the package version can provide significantly higher coverage, i.e., an extra 50\% improvement compared to tests in the latest commit. 
In particular, the number of packages with coverage $\geq$ 80\%, $\geq$ 20\%, and $>$ 0\% increases from 101 to 154, from 139 to 203, and from 163 to 218, respectively. 
Finally, curve \emph{original} shows that the strategy used by \fw{} to resolve differences between releases in the \npmreg{} and the selected revisions leads to 6.5\%, 10.8\%, and 11\% more packages with coverage $\geq$ 80\%, $\geq$ 20\%, and $>$ 0\%, respectively.

Overall, our evaluation shows that \fw{} can locate original tests and automatically measure code coverage for about half of the packages in the \ts{} without any manual intervention. 
Unfortunately, no original tests can be found for the other half of the \ts{}.
We discuss the reasons in Section~\ref{sec:discussion}.

\subsection{Dependent Tests}\label{sec:eval_dep}

\begin{figure}[t]
    \centering
        \includegraphics[width=0.5\textwidth]{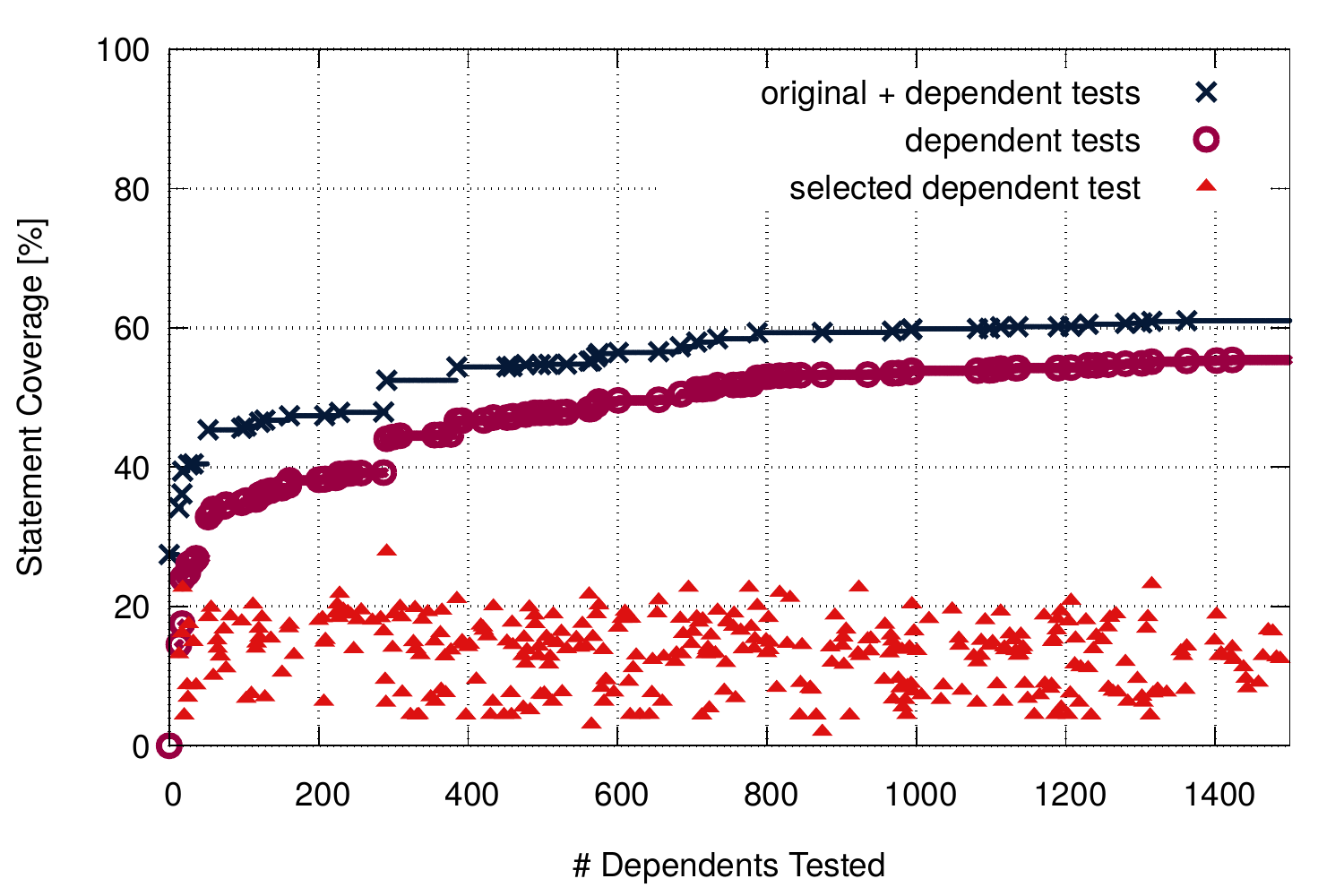}

    \caption{Code coverage for {\tt lodash}. Improvement to statement coverage wrt.~the number of dependents tested by \fw{}. No compaction of the test suite is done. }
        \label{fig:trace}
\end{figure}

\begin{figure*}[t]
    \centering
    \begin{tabular}{M{0.3\textwidth}M{0.3\textwidth}M{0.3\textwidth}}
         \hspace{-0.04\textwidth}   \includegraphics[width=0.3\textwidth]{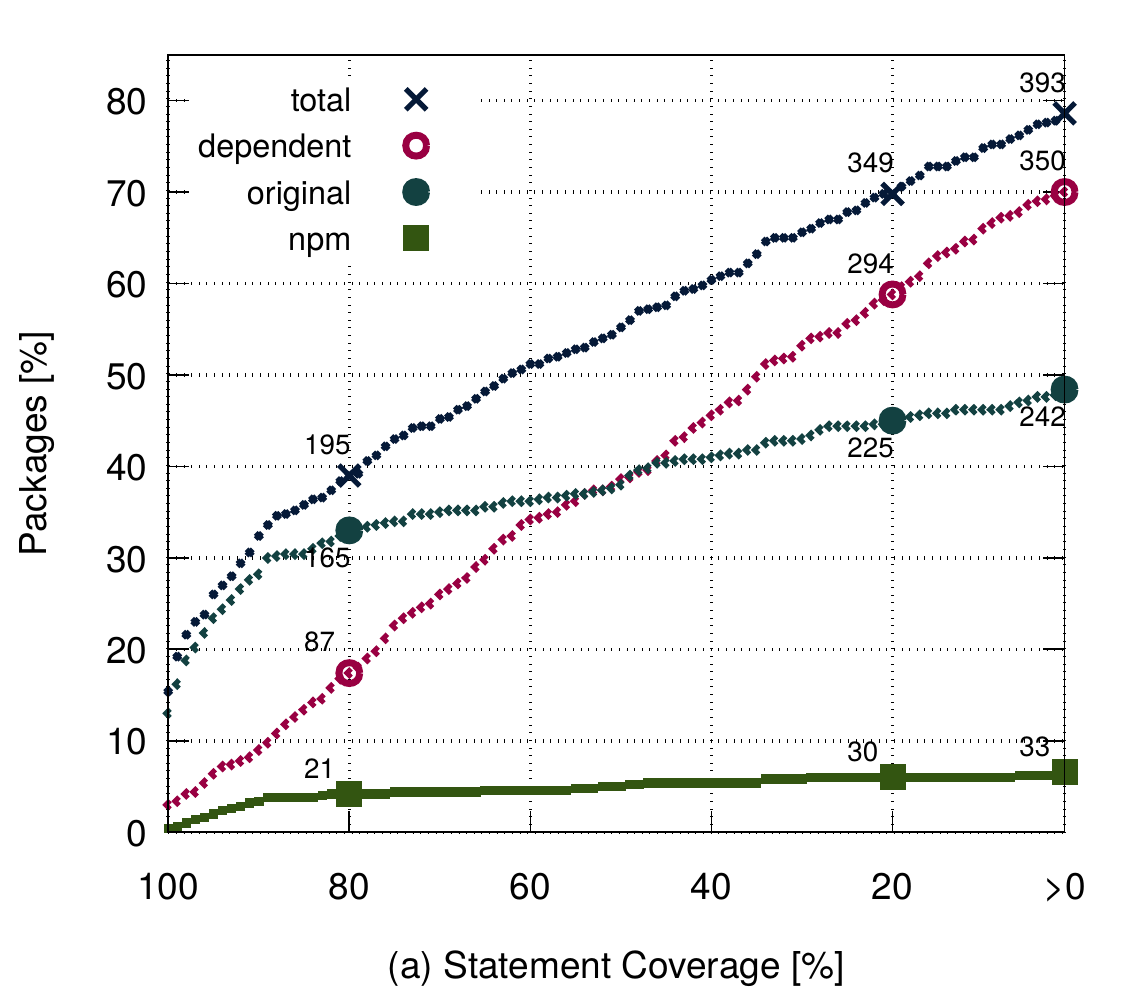} &
   \hspace{-0.04\textwidth}         \includegraphics[width=0.3\textwidth]{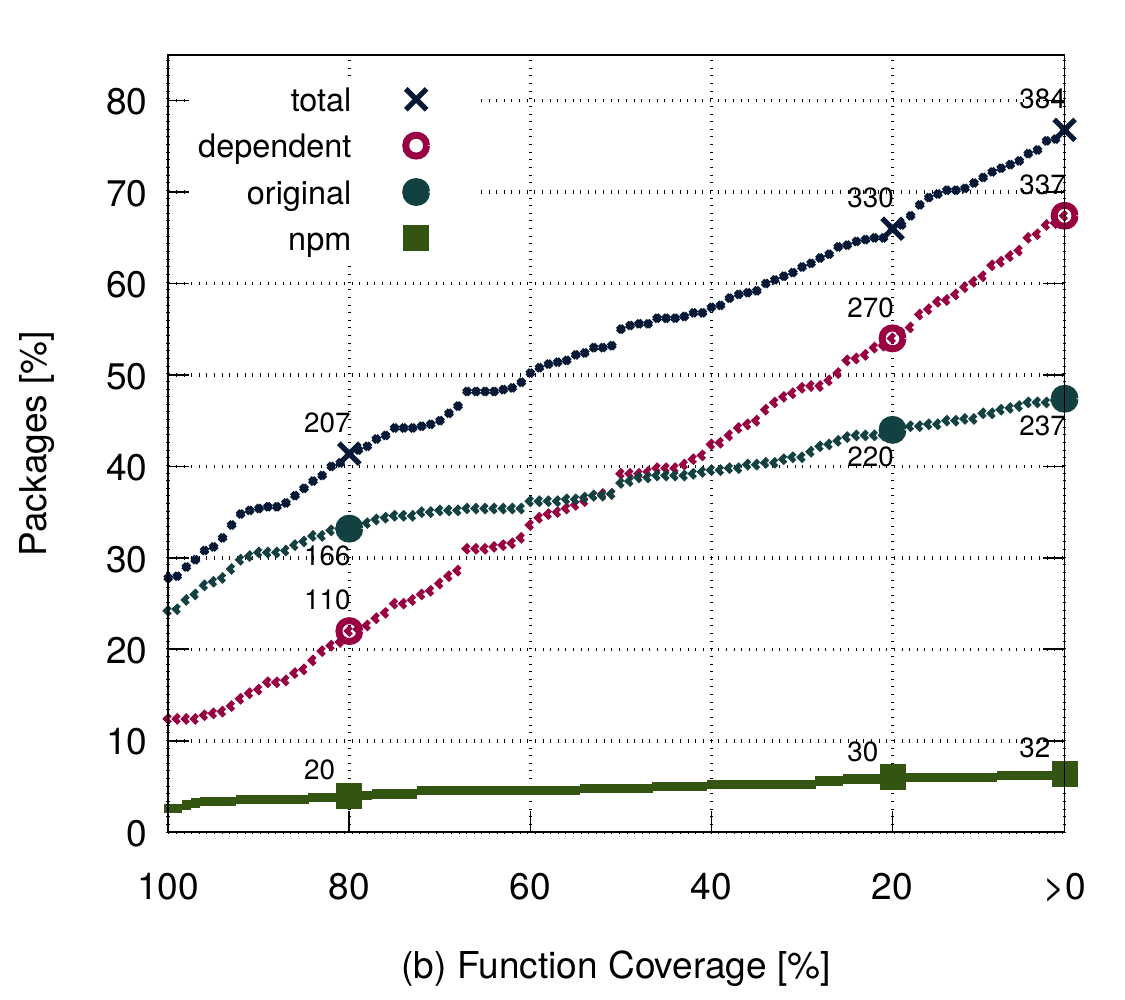} &
   \hspace{-0.04\textwidth}         \includegraphics[width=0.3\textwidth]{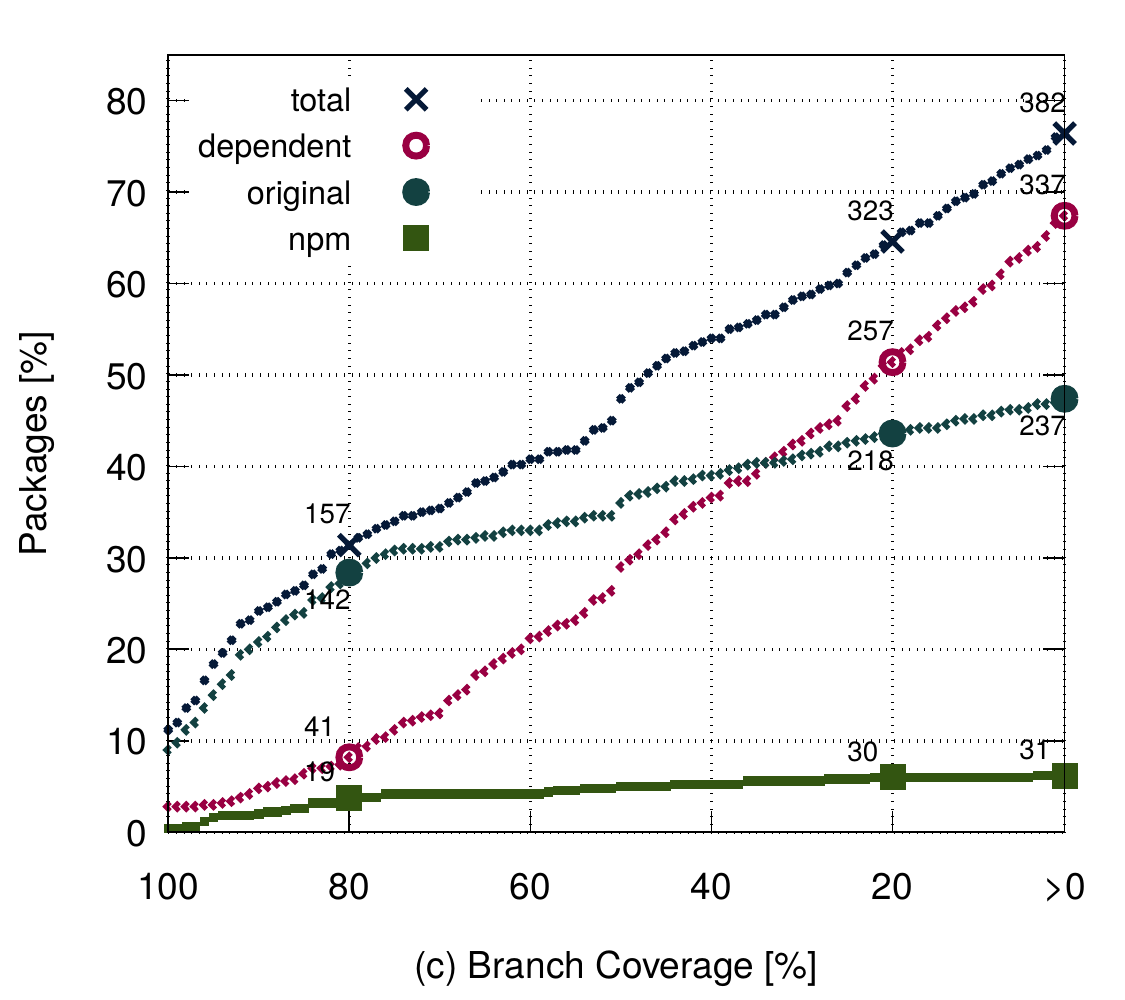} \\
    \end{tabular}
    \caption{Code coverage of the test suite generated by \fw{}. Original vs.~dependent tests. Points in the curve represent the number of packages with coverage $\geq$ 80\%, $\geq$ 20\%, and $>$ 0\%, respectively.  The evaluation set is composed of \TOTAL{} packages. }
    \label{fig:all-improve}
\end{figure*}

In this section, we evaluate the coverage resulting from dependent packages. 
First, we show how \fw{} discovers dependent tests focusing on one very popular package called \code{lodash}.
Then, we show the overall code-coverage improvement for the whole \ts{} when including dependent tests.

\subsubsection{Discovering Dependent Tests for lodash}
\code{lodash} is one of the most popular Node.js packages.
The package has 44.6K GitHub stars, and it was downloaded more than 1.18 billion times in 2019.
This subsection focuses on the discovery of dependent tests for \code{lodash}.
In particular, we consider version \code{4.17.15} of the package (last release in 2019). 

Figure~\ref{fig:trace} shows the gradual code coverage improvement for \code{lodash} in relation to the number of dependent tests found by \fw{}, considering the first 1500 dependent packages tested (selection criteria described in Section~\ref{sec:deptests}).
The x-axis shows the number of dependents that have been tested, while the y-axis represents the statement coverage.
The scattered points (labeled ``selected dependent test'') represent individual dependent tests 
to be included in the test suite.
The curve in the middle (labeled ``dependent tests'') represents the cumulative coverage of the test suite considering only the dependent tests. 
The curve at the top (labeled ``original + dependent tests'') represents the total code coverage achieved considering both the original tests and the dependent tests. 
The left end of this curve has a value of 27.5\%, which is the coverage of the original tests 
for \code{lodash@4.17.15}. 
Every increase in the value of the two curves is highlighted with a dot that corresponds to the dependent test that leads to the increase.
Overall, 388 valid dependent tests
(i.e., finishing successfully with non-zero code coverage) are found for \code{lodash}
among the first 1500 dependent packages, as highlighted by the scattered points.

In the figure, we observe that the final coverage of the dependent tests (the right end of the ``dependent tests'' curve) is 55.4\%, 
which is significantly higher than the coverage of the original tests (27.5\%, i.e., the starting value of the top curve). 
This shows that for \code{lodash}, the dependent tests found by \fw{} feature an even higher statement coverage than the original tests. 
At the same time, we observe that there is a 10.6\% difference between the right ends of the two curves.
This gap represents the presence of a portion of code that is only covered by the original tests of \code{lodash}. 

By inspecting the differences in the coverage reports generated by \fw{}, we find that the main reason why \code{lodash} has a lower coverage for original tests in comparison to dependent tests is that the original tests focus only on the main code file of the package (\code{lodash.js}) without testing many other small utility files (there are overall 1046 JavaScript source-code files in \code{lodash@4.17.15}).
However, such files are used by the dependents and covered by some dependent tests. 
On the other hand, the original tests of \code{lodash} still cover some code that is not used by any of the 1500 dependent tests.
To narrow the difference, more dependents have to be tested.
While doing so is time-consuming, 
such a search process has a one-time cost---once the test suite has been assembled, we do not need to repeat the search.

In summary, our evaluation shows that both sources of tests (original and dependent) complement each other, leading to a test suite with higher code coverage. 

\subsubsection{Code Coverage Including Dependent Tests}

Figure~\ref{fig:all-improve} compares the coverage resulting from dependent tests (``dependent''
curve) with the one resulting from the tests included in  the \npmreg{} (``npm'' curve),  from original tests (``original'' curve), as well as with the total coverage of the test suite assembled by \fw{} (``total'' curve). 
Similarly to Figure~\ref{fig:ori-improve}, each of the three subfigures focuses on a different code-coverage metric.
Again, we explain our results focusing only on statement coverage, shown in Figure~\ref{fig:all-improve}a.

By looking at the right end of the curves, we can observe that more packages are found with dependent tests (350) than with original tests (242).
This can be explained by the fact that the possibility of finding original tests significantly depends on the existence of correctly configured and runnable tests in the development repository of the target module.
The presence of many dependents increases the chances of locating correctly configured tests, and hence valid dependent tests for a module. 
 
We observe that 165 out of the 242 packages with valid original tests have at least 80\% code coverage, while only 87 out of 350 packages with valid dependent tests have a code coverage $\geq$ 80\%. 
This shows that original tests from the maintainers alone tend to have better coverage compared to only dependent tests. 
This can be explained as some functionalities or corner cases (such as exception handling) of the target package are rarely or never exercises by dependents, but are nonetheless tested by the package maintainers. 
When including dependent tests in the suite assembled by \fw{}, the number of packages with at least 20\% coverage is increased by 50\% (from 225 to 349), while the number of packages with 80\% or more coverage is increased by almost 20\% (from 165 to 195).
This indicates that the dependent tests are not subsumed by the original tests, and highlights the fundamental role of dependent tests in increasing code coverage. 

Overall, the test suite assembled by \fw{} includes both original and dependent tests, written by different groups of developers (i.e., the maintainers of a package and the developers using it).
Each kind of test can cover code portions that are not covered by the other, thus complementing each other.
\fw{} leverages this synergy and exploits both sources of tests, which results in a more complete test suite with increased code coverage.


\begin{table*}[t]
    \center 
\begin{tabular}{ c | c c | c c | c | c}

\multirow{2}{*}{\textbf{Package(s)}} &
  \multicolumn{2}{c|}{\textbf{\# Tests Included}} &
  \multicolumn{2}{c|}{\textbf{Execution Time}} &
  \multirow{2}{*}{\textbf{\begin{tabular}[c]{@{}c@{}}Comp. \\Rate\end{tabular}}} &
  \multirow{2}{*}{\textbf{\begin{tabular}[c]{@{}c@{}}Comp.\\Time\end{tabular}}}\\ \cline{2-5}
        & \textbf{Before} & \textbf{After} & \textbf{Before} & \textbf{After} &      &   \\ \hline
    \multirow{2}{*}{\code{\footnotesize lodash}}  & 1 original             & 1 original              &     \multirow{2}{*}{1h28m}          &     \multirow{2}{*}{19m29s}            &     \multirow{2}{*}{4.5} &     \multirow{2}{*}{40s} \\
  & 388 dependent             & 39 dependent              &           &             & & \\ \hline
    all     & \multirow{2}{*}{9.82}              & \multirow{2}{*}{1.70}                   & \multirow{2}{*}{2m10s}          & \multirow{2}{*}{22s}   & \multirow{2}{*}{6.0}    & \multirow{2}{*}{0.13s} \\
        (average)     &               &                    &          &   &     &  \\ 
\end{tabular}
\vspace{2mm}
    \caption{Evaluation of the test-suite compaction algorithm applied by \fw{}. Columns ``Before'' and ``After'' report metrics measured on the test suite before and after running the compaction algorithm, respectively. The evaluation set is composed of \TOTAL{} packages. }
\label{table:compact}
\end{table*}

\subsection{Test-suite Compaction}\label{eval:compact}
Here, we evaluate the algorithm used by \fw{} to compact the final test suite (see Section~\ref{sec:compact}). 
This evaluation has been conducted on a machine running Ubuntu Server version 18.04 (kernel version 4.15.0-66), equipped with 128~GB of RAM and 8 CPU cores (2.7~GHz).
Table~\ref{table:compact} compares the test suite before and after compaction.
The first row shows results for the \code{lodash} package, while the second row reports the average-per-package results for running the algorithm on the entire \ts{}.

The table compares the number of tests included in the suite and the total test execution time, before and after the compaction.
The last two columns report the compaction rate and the time spent running the compaction algorithm. 

The compaction algorithm chooses for \code{lodash} the single original test and 39 additional dependent tests, out of the 388 valid dependent tests for this package, which reduces the time needed to execute all tests from 1~hour and 28~minutes to 19~minutes and 29~seconds, with a compaction rate of 4.5.
The execution time of the compaction algorithm itself is 40 seconds, including reading and processing the coverage reports collected by \fw{}.
When considering all the test suites generated by \fw{}, the last row shows that, on average, our compaction algorithm takes very little time (0.13s) to finish and yields a compaction rate of 6.0.


\subsection{Case Study: Running DPA with Extended Code Coverage}\label{sec:eval_dpa}
As the JavaScript language is highly dynamic, 
developers often rely on DPA tools to better understand the runtime properties of a JavaScript application. 
To be effective, it is crucial that DPA is applied to workloads covering a significant portion of application code, as DPA cannot provide any information on code that it is not executed. 
Previous work has shown~\cite{m-autobench,m-ecoop-2019} that unit tests are a viable source of workloads for discovering runtime properties with DPA. 
In this section, we show that the test suite provided by \fw{} can also be used to extend the effectiveness of DPA tools.
In particular, we show that existing DPAs can provide more information and locate more performance problems if executed on the test suite provided by \fw{} than simply relying on the tests included in the \npmreg{} or on the original tests of a package.

To run DPA tools, \fw{} integrates NodeProf~\cite{m-cc},
a DPA framework for Node.js, which offers support for the latest ECMAScript features. 
Thanks to this integration, any existing or new DPA developed with NodeProf can be automatically applied to the test suite assembled by \fw{}.

In this section, we focus on 
 two existing DPAs (\code{non-contiguous-array} and \code{typed-array}) included in JITprof~\cite{jitprof2015}, a collection of DPAs that can run on NodeProf to identify a variety of code patterns that may jeopardize optimizations performed by the just-in-time (JIT) compiler of the JavaScript engine.
 As output, both analyses produce a set of source-code locations, each representing one finding.
 We provide more details on the two DPAs below. %

 \begin{compactitem}
    \item \code{non-contiguous-array} detects array writes at indices that are outside the array boundaries. In such cases, the array will be automatically extended, and the gap between the old array boundary and the write index will be filled with \code{undefined} objects. Such a pattern can degrade performance, 
        as the JIT compiler may not be able to treat the array as a primitive one that can be stored as a sequential memory buffer.
\item \code{typed-array} locates arrays that store only primitive types (such as \code{integer} or \code{byte}) and are accessed frequently (the threshold used in the analysis is 1000 array accesses). These arrays could be replaced by specific typed array-like objects in JavaScript (such as \code{Int32Array} or \code{Uint8Array}), allowing faster read and write accesses. 
\end{compactitem}

We apply the two DPAs on the following test suites, obtained by \fw{} from the \TOTAL{} packages in the \ts{}:
1) tests contained in the package release on the \npmreg, 
2) original tests, 
3) the final test suite including both original and dependent tests. 
We report the findings (only for the code within the target package) reported by the two DPAs in Table~\ref{table:dpafindings}.
The left part of the table shows the number of findings reported by the DPAs executed on the different test suites, while the right part of the table reports the number of packages with at least one finding. 

From the table, we can observe that with the inclusion of more tests, running the analyses results in more reported findings.
Compared to simply running the tests found in the \npmreg{} for the \ts{},  
by executing the DPAs on the original tests from the development repository the number of findings for {\tt non-contiguous-array} and {\tt typed-array} increases from 5 to 8 and from 36 to 97, respectively, while the number of packages with at least one such finding increases from 3 to 6, and from 6 to 21, respectively. 
Executing the DPAs on the complete test suite generated by \fw{} (including both original and dependent tests) further increases the number of findings reported by the two analyses, from 8 to 20 ({\tt non-contiguous-array}), and from 97 to 201 ({\tt typed-array}).
Moreover, the number of packages where a performance problem is found increases from  6 to 11 ({\tt non-contiguous-array}) and from 21 to 40 ({\tt typed-array}).

Overall, our evaluation results show that the test suite generated by \fw{} can help DPA tools yield additional important results, thanks to the extended test-coverage provided by dependent tests. 

\begin{table}[t]
\center \setlength{\tabcolsep}{4pt}
\begin{tabular}{ c | ccc | ccc}
\multirow{2}{*}{\bf DPA}                      & \multicolumn{3}{c|}{\bf \# Findings}   & \multicolumn{3}{c}{\bf \# Packages } \\ \cline{2-7}
                 & npm & original & final & npm    & original    & final   \\ \hline
\begin{tabular}[c]{@{}c@{}}{\code{\footnotesize non-contiguous-}}\\  \code{\footnotesize array}\end{tabular} & 5 & 8       &  20     & 3      & 6          & 11        \\ \hline
\code{\footnotesize typed-array}          & 36 & 97     & 201      & 6     & 21         &    40     \\ 
\end{tabular}
\vspace{2mm}
    \caption{Applying DPAs to the test suite generated by \fw{}. Number of findings and number of packages where at least one finding is reported.}
    \label{table:dpafindings}
\end{table}


\section{Lessons Learned}\label{sec:lessons}

In this section, we discuss the lessons learned during the development and evaluation of \fw{}. 

\indent {\bf {\it a)} Dependent tests are useful.}
Dependent tests are a very valuable source of tests besides the original tests.
First, dependent tests alone can cover a good portion of the code of a package (e.g., for about 35\% of the packages considered in our evaluation, dependent tests can achieve more than 60\% statement coverage). 
Second, dependent tests represent a valid alternative to original tests to assess package quality, especially for packages that do not include correctly configured and runnable tests (i.e., more than half of the packages in our evaluation).
In addition, dependent tests can be used to study the impact of newly found bugs~\cite{10.1145/3377811.3380442} and the backward compatibility~\cite{10.1145/3377811.3380436} of the target package.
Finally, dependent tests can significantly increase the effectiveness of DPA tools, which can analyze a higher portion of code when compared with running only the original tests. 
Overall, original and dependent tests can cover code portions that are not covered by the other, thus complementing each other. This remarks the importance of considering both kinds of tests when evaluating package quality, as done by \fw{}.

\indent {\bf {\it b)} Package releases in the npm registry often exclude tests that can only be found in their development repository.} As a result, testing a \emph{specific} package release in the npm registry without knowledge about how versions and tests are organized in the development repository is difficult. 

\fw{} simplifies this task by automatically selecting the right tests for a specific package version. This process is transparent to the user and does not require extra effort in manually investigating the development repository.

\indent {\bf {\it c)} Many packages, including popular ones, lack the necessary configuration for their tests}. The npm registry does not require a package to have valid tests, making it harder to standardize package testing. In our study, we find that often 
packages require the installation of extra packages (e.g., testing harnesses like \code{mocha}), package managers (e.g., \code{yarn}), or system libraries (e.g., \code{curl}) to run their tests. However, such packages and libraries are neither specified as dependencies by packages' developers, nor installed by default in a standard testing machine. As a result, even after installing dependencies via the standard \code{npm install} command, the tests will fail. 
This might be explained by the fact that package maintainers have these prerequisites configured in their own testing environment, and thus do not notice that such dependencies would be missing for other developers. 
As not all the missing prerequisites are well-known, it is not possible to preinstall all the prerequisites that work for all packages.

\fw{} is designed to handle this case. First of all, \fw{} preinstalls about 30 popular prerequisites by statically analyzing the testing scripts of the target packages. In addition, when a test fails, \fw{} analyzes the error output to determine whether the failure is caused by missing dependencies or runnable commands in the environment and tries to install them automatically. Only if this fails, \fw{} notifies the user and asks for manual inspection.
The error handling applied by \fw{} can greatly reduce test failures due to misconfigurations and allows finding more valid original and dependent tests when assembling the test suite.

\indent {\bf {\it d)} Always sandbox the tests.}
As some packages may include scripts that cause side-effects on the host machine (e.g., the installation of additional tools or the forced setting of global configurations) or can even be malicious, it is fundamental to execute tests in a sandboxed environment. \fw{}  runs all package tests in Docker containers to prevent such side-effects.

\indent {\bf {\it e)} Scalability matters.}
As a package could have thousands of dependents (e.g., \code{lodash} has more than 120K dependents), searching for all suitable dependent tests may take a very long time. It is therefore very important to implement a scalable strategy that can speed up the assembly of the test suite.
In \fw{}, the search for dependent tests can run in parallel, evenly distributing test lookup among workers running in different Docker containers. 


\section{Discussion}\label{sec:discussion}

In this section, we discuss important aspects of our approach, while also introducing ideas for future work and the limitations of our technique. %

\indent {\bf {\it a)} Coverage metrics. }
\fw{} currently employs three coverage metrics (i.e., statement, function, and branch coverage) to measure test quality, relying on \code{istanbul.js}. 
Such metrics are widely adopted and have been used in many other studies on JavaScript applications~\cite{7927978,npmminer,ivankovic2019code}.
Nevertheless, we plan to add support for more coverage metrics, such as modified condition/decision coverage (MC/DC)~\cite{chilenski1994applicability,ivankovic2019code}. 
As MC/DC is not subsumed by the existing metrics in \fw{}, and there is no tool 
focusing on such a metric for JavaScript, some original tests with high statement, function, or branch coverage may fall short in MC/DC coverage, as developers currently cannot measure such a metric on their tests. 
The test suite assembled by \fw{} may improve such a metric as well, thanks to the provision of numerous dependent tests. 

\indent {\bf {\it b)} Test oracles.}
Having a good test oracle~\cite{6963470} can help better distinguish correct program behaviors from incorrect ones.
Currently, the tests assembled by \fw{} are based on a package-level granularity, and whether an original or dependent test runs successfully is validated by looking at the exit code of the \code{npm test} command, which is the only general way of telling whether a test for a JavaScript package succeeds or not.
However, such an oracle is rather coarse-grained, as a package test can contain many individual test cases.
In the future, we plan to recognize individual test cases for the most popular testing harnesses, and integrate some oracle-generation techniques~\cite{6963470,6693121,10.1145/3266237.3266273,survey2017} to achieve a more fine-grained test selection.

\indent {\bf {\it c)} Failed tests.}
Although failed tests are not included in our automatically assembled test suite, it is still worthwhile to understand the failure causes. 
First, \fw{} stores the standard error output of a test, where some general causes of failure can be detected, such as missing dependencies.
In addition, \fw{} measures coverage for the failed tests, which can be useful to indicate the failure cause. 
For example, if a test is found with zero coverage, the test has usually failed in an early stage before any code of the package is executed.
A non-zero coverage  (meaning that a part of the source code has been executed, but the test still fails) is usually caused by the failure of at least one test case in the package test.
In our evaluation, original tests of 205 packages in the \ts{} fail with zero coverage, while original tests of 55 other packages fail with non-zero coverage.
We plan to integrate a more in-depth analysis to automatically identify more accurate failure causes of a package.

\indent {\bf {\it d)} Determinism of the test suite.}
A deterministic test should either always pass or always fail for the same code under testing.
Such a property is a demanded feature for software testing, especially for regression testing~\cite{10.1145/2635868.2635920}. 
However, how to detect and fix tests with non-deterministic outcomes (often called ``flaky tests'')~\cite{10.1145/3377811.3381749} remains an interesting and challenging research question~\cite{10.1145/2635868.2635920,8453104}.
In \fw{}, flaky tests are detected by running the test (without coverage measurement) multiple times, and are not included in the test suite.

Another important property is deterministic code coverage, i.e., repeated executions of a test cover the same part of the package code.
We plan to track this property by re-running the test-coverage measurements multiple times. 
If the covered program elements differ in repeated measurements, we can conservatively measure the code coverage of a package by taking the intersection of the covered program elements in different runs, and compute our test-coverage metrics from the intersection.

\indent {\bf {\it e)} Node.js version.}
\fw{} is developed using Node.js.
The version used in our evaluation is Node~12 (the latest LTS release at the time of writing).
We noticed that some old packages have compatibility issues with Node 12.
Moreover, some of the most recent packages use the latest Node~14 features (e.g., \code{ES module}, which is no longer experimental from Node~14) that are still not fully supported by Node~12.
As future work, we plan to extend our evaluation to also consider package compatibility with the latest Node.js version and with other LTS releases. 
We also plan to allow users to filter tests (not) supporting specific Node versions. 

\indent {\bf {\it f)} Code transformations.}
Some packages are tested together with code transformation tools that can change the source code being tested. 
For example, some packages rely on linting tools such as \code{eslint} for formatting, some others rely on a code instrumentation framework such as \code{babel} to parse and generate code, and yet other packages are developed using TypeScript~\cite{typescript} and then compiled into JavaScript code using a TypeScript transpiler. 
Using such tools can lead to different versions for some source-code files, possibly decreasing the code coverage computed by \fw{}.

We plan to improve \fw{} by detecting and disabling unnecessary code transformations.
In cases where this is impossible, we will ensure that \code{istanbul.js} has the highest priority in code instrumentation compared to other tools, such that coverage reports can be generated before code transformations take place.

\indent {\bf {\it g)} Uniform coverage tools.} 
\fw{} uses the popular \code{istanbul.js} to measure the code coverage of a package. 
When evaluating and comparing the coverage of different packages in the npm registry, it is important to use the same version of \code{istanbul.js} for all target packages in order to generate compatible coverage reports. However, many packages integrate different versions of \code{istanbul.js} in their tests; unfortunately, different versions generate coverage reports that are not compatible with each other. \fw{} mitigates this problem by unifying the version of \code{istanbul.js} used, so that all coverage reports obtained on the same code are compatible and can be compared. \fw{} currently does not consider cases where the testing harness itself (such as \code{jest}) has a different version of \code{istanbul.js} integrated. We plan to deal with such cases in the future release of \fw{}.  


\section{Related Work}\label{sec:rw}

Code coverage is an important metric for assessing the quality of code testing~\cite{ivankovic2019code,7927978,survey2017,10.1145/3180155.3180209}. 
A study~\cite{7927978} on the code coverage of JavaScript projects highlights that many of the considered JavaScript applications lack tests or suffer from limited code coverage. 
In such a study, the measurement of code coverage relies on a partially manual workflow.
Thus, such an approach does not scale to a large number of packages, contrary to our approach.
In addition, the study does not propose any solution to increase code coverage, as we do. 

There are many studies~\cite{10.1145/3408302,zimmermann2019small,10.1145/3180155.3180209,10.1145/2901739.2901743,npmminer} based on data mining and static analysis of the npm ecosystem. A study by Zimmermann et al.~\cite{zimmermann2019small}, based on the analysis of the packages' dependencies in the \npmreg{}, shows that the npm ecosystem faces single-point-of-failure risks, as problems in individual packages may impair large parts of the entire ecosystem.
Another study by Trockman et al.~\cite{10.1145/3180155.3180209} analyzes repository badges of npm packages to understand the quality of the npm ecosystem, finding a positive correlation between the assignment of a coverage badge and the presence of more test code. %
The study from Hejderup et al.~\cite{10.1145/3183399.3183417} analyzes the dependency among different versions of npm packages by constructing the call graph at the function level using static analysis.
NpmMiner~\cite{npmminer} employs static analysis on 2000 packages to highlight metrics such as cyclomatic complexity~\cite{gill1991cyclomatic} and lines of code, as well as pinpoint code linting issues. 
All these studies are based on information that can be obtained without executing the test code. 
However, despite the easy applicability of such approaches based on static information, these approaches fall short in understanding the dynamic behavior of the packages. 
It is also well known that it is difficult to achieve sound whole-program static analysis for JavaScript applications~\cite{survey2017,TAJS,taser2020,Andreasen:2017:SAI:3088515.3088521} due to the dynamic nature of JavaScript.

The literature is rich in work proposing test generation techniques aiming at generating
new tests by exploring the input space and event space of JavaScript applications to improve code coverage.
Some test-generation tools~\cite{Mesbah:2012:CAW:2109205.2109208,7372008} mainly focus on client-side web applications, increasing code coverage by exploring browser states (e.g., DOM changes related to mouse and keyboard events) and cannot be applied to server-side packages that typically use dedicated APIs (e.g., for performing I/O operations on the network or file system). 
Many test generation tools still rely on the input of existing tests. 
Artemis~\cite{6032496} uses feedback-directed testing to generate test inputs by observing the effects of inputs on existing tests.
Testilizer~\cite{10.1145/2642937.2642991} generates tests based on the input data, event sequences, and assertions learned from the existing tests. 
Our approach can be used to provide the valuable initial input seeds for such techniques,  as our tests come from existing valid tests.

Mutation testing~\cite{PAPADAKIS2019275} is another effective way to evaluate the quality of existing tests by measuring the percentage of mutants that they kill. 
Well-known mutation testing tools for JavaScript include Mutode~\cite{10.1145/3213846.3229504} and MUTANDIS~\cite{6569719}. 
In the future, such tools could be integrated with \fw{} to further improve the quality of the assembled test suite.

Fuzzing~\cite{8371326,jsfuzz,10.1145/3360600} is another technique that generates new test cases by providing new random inputs to a program.
Unfortunately, existing fuzzing tools for JavaScript are  limited due to the dynamic typing used in JavaScript. 
JSFuzz~\cite{jsfuzz} is one of the few fuzzing tools available for JavaScript. JSFuzz takes a \textit{fuzz target} which calls the function or library to be tested and then generates new inputs in an infinite loop using a coverage-guided algorithm. The approach is not fully automatic as the fuzz target has to be provided manually. 
In addition, it does not work for functions that take dynamic objects as inputs.

In general, due to the dynamic nature and event-driven programming model of JavaScript,
fully exploring the input and event space of JavaScript applications automatically without application-specific knowledge remains a challenging topic~\cite{survey2017}.
Our approach of using original tests as well as tests from the package's dependents makes use of meaningful inputs from the actual package developers and can be complementary to such automatic test generation techniques.

To deal with the limitations of static analysis for JavaScript applications, many dynamic analysis frameworks~\cite{m-cc,jalangi13} and tools~\cite{m-cgo,jitprof2015,meminsight2015,m-ecoop-2019,Alimadadi:2018:FBP:3288538.3276532} have been proposed. 
One known issue of dynamic analysis is that it can only analyze the code that is executed.
We have shown that the test suite assembled by \fw{} can benefit DPA tools, since an increased code coverage can expand the scope of DPA tools. 


\section{Conclusion}\label{sec:conclusion}
In this paper, 
we introduce \fw{}, a new framework that automatically assembles a test suite for packages hosted in the~\npmreg. 
The suite generated for each target package includes the original tests written for the target package, as well as dependent tests written for its dependent packages. 
\fw{} uses the original tests to assess the code coverage of packages in the~\npmreg{}, and exploits dependent tests to increase the code coverage of the target packages. 
\fw{} is fully automatic and can scale to a large number of packages available in the~\npmreg{}.
Our evaluation results show that \fw{} can assess code coverage for numerous popular packages, while also extending their code coverage. 
Moreover, our results demonstrate that the test suite assembled by \fw{} can increase the effectiveness of existing DPAs, allowing them to locate performance problems that cannot be identified with the original tests of a package. 

\section*{Acknowledgement}\label{sec:ack}

This work has been supported by Oracle (ERO project~1332), by the Hasler Foundation (project~20022) and by the Swiss National
Science Foundation (project 200020\_188688).

\bibliography{bibexport}

\end{document}